\documentclass{article}
\usepackage{spconf,amsmath,graphicx}
\usepackage[caption=false,font=footnotesize]{subfig}

\title{LSTM-TDNN with convolutional front-end for Dialect Identification in the 2019 Multi-Genre Broadcast Challenge}
\name{Xiaoxiao Miao$^{1,2,3}$, Ian McLoughlin$^{4,1}$
\thanks{Thanks to China Scholarship Council for funding to conduct this research at the University of Kent, Medway, UK. This work is also partially supported by the National Key Research and Development Program (Nos. 2016YFB0801203, 2016YFB0801200) , the National Natural Science Foundation of China (Nos. 11590774, 11590770), the Key Science and Technology Project of the Xinjiang Uygur Autonomous Region (No.2016A03007-1)}
}
\address{
$^1$School of Computing, The University of Kent, Medway, UK\\
$^2$Key Laboratory of Speech Acoustics and Content Understanding, \\ Institute of Acoustics, Chinese Academy of Sciences\\
$^3$University of Chinese Academy of Sciences\\
$^4$The National Engineering Laboratory for Speech and Language Information Processing,\\The University of Science and Technology of China}

\begin{document}
\maketitle
\begin{abstract}
This paper presents a novel Dialect Identification (DID) system developed for the Fifth Edition of the Multi-Genre Broadcast challenge, the task of Fine-grained Arabic Dialect Identification (MGB-5 ADI Challenge). The system improves upon traditional DNN x-vector performance by employing a Convolutional and Long Short Term Memory-Recurrent (CLSTM) architecture to combine the benefits of a convolutional neural network front-end for feature extraction and a back-end recurrent neural to capture longer temporal dependencies. Furthermore we investigate intensive augmentation of one low resource dialect in the highly unbalanced training set using time-scale modification (TSM). This converts an utterance to several time-stretched or time-compressed versions, subsequently used to train the CLSTM system without using any other corpus. In this paper, we also investigate speech augmentation using MUSAN and the RIR datasets to increase the quantity and diversity of the existing training data in the normal way. Results show firstly that the CLSTM architecture outperforms a traditional DNN x-vector implementation. Secondly, adopting TSM-based speed perturbation yields a small performance improvement for the unbalanced data, finally that traditional data augmentation techniques yield further benefit, in line with evidence from related speaker and language recognition tasks.
Our system achieved 2nd place ranking out of 15 entries in the MGB-5 ADI challenge, presented at ASRU\,2019.
\end{abstract}
%

\section{Introduction}
\label{sec:intro}
Dialect Identification (DID), the task of automatically identifying which dialect an utterance contains from among a set of candidate dialects, is a task closely related to language identification (LID).
However similar dialects tend to be much more closely spaced in feature similarity than for LID.
Both LID and DID are strongly affected by issues such as speaker variance, background noise, channel mismatch, vocabulary deficiency, duration mismatch and so on, and both tend to become more difficult as the number of languages or dialects included in the set is increased.
However both are increasingly popular research targets with important real-world applications.
For example, Chinese people from different regions speak their own dialects even though they share the same written text. 
The acoustic characteristics between dialects can be substantially different in some cases, but in others are extremely subtle.
For this reason, dialects are generally harder to distinguish than languages, and suffer more from overlap in the their acoustic, linguistic and speaker characteristics. 
Effective DID is potentially beneficial for improving automatic speech recognition (ASR) systems, since dialect-specific information allows language models to be optimized.

The Multi-Genre Broadcast (MGB) challenge committee established an Arabic dialect dataset in 2016, and holds a DID challenge series, which focuses on closely-spaced Arabic dialects. 
MGB-5 ADI in the summer of 2019 is the latest challenge, and this paper present the authors' submission to that challenge. Results were announced at the ASRU 2019 workshop, held in Singapore in December 2019.

\begin{figure*}[thb!]
\centering
\subfloat[][TDNN]{\includegraphics[height=3cm]{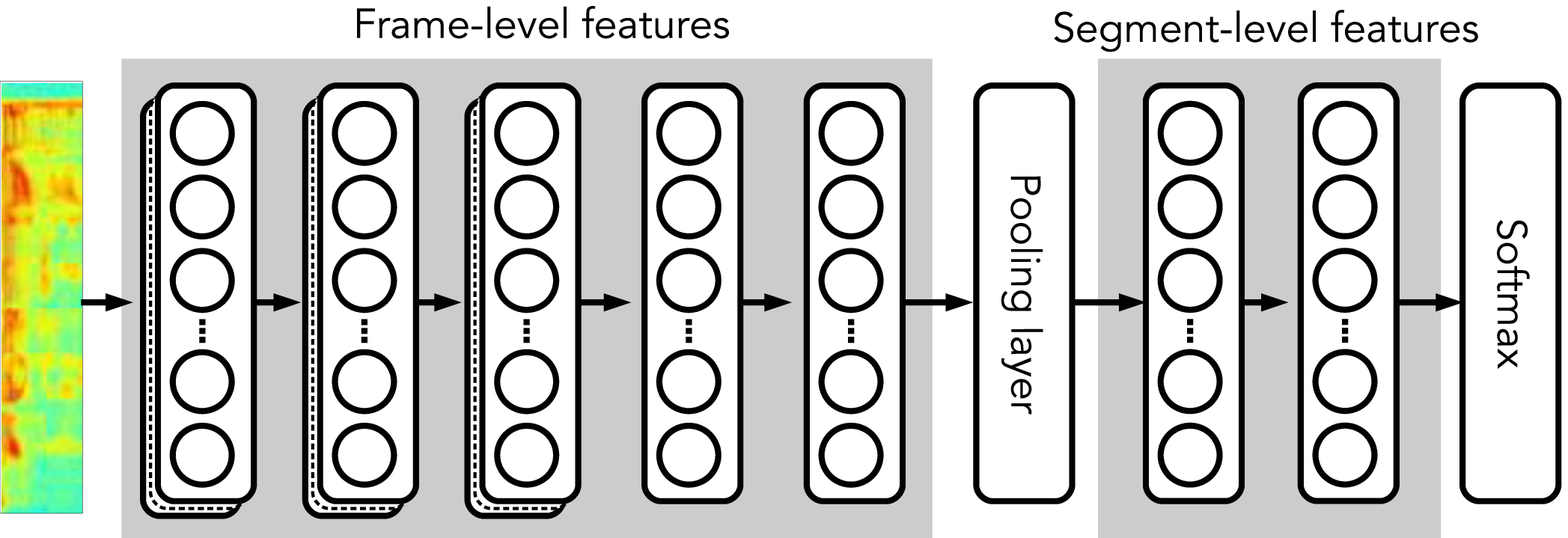}}\\
\subfloat[][CNN-LSTM-TDNN]{\includegraphics[height=3cm]{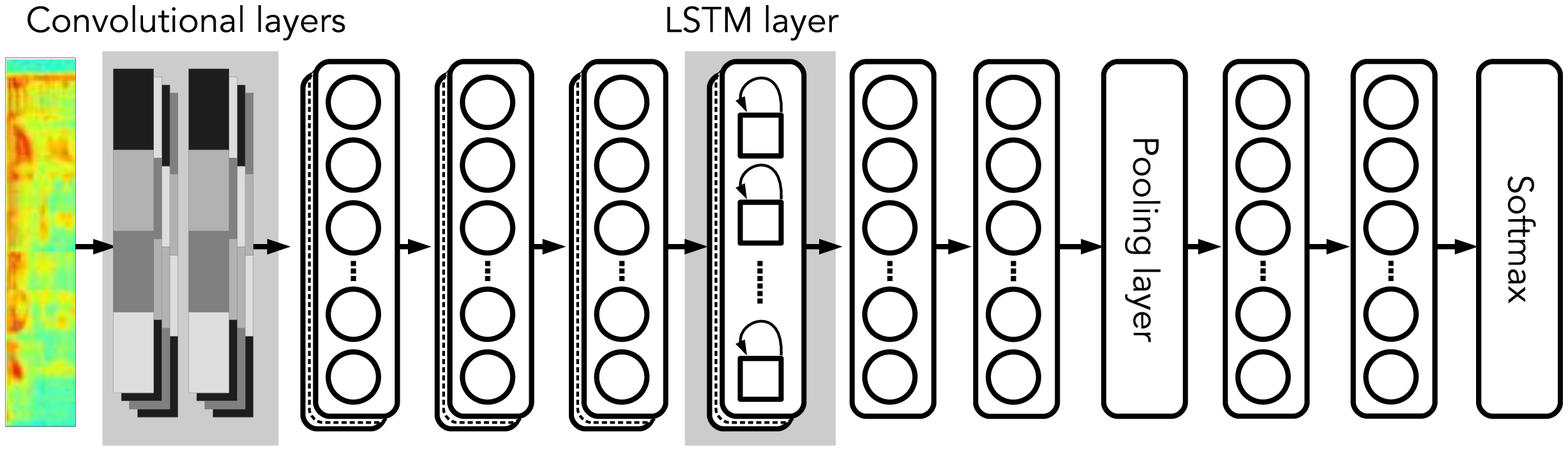}}\vspace{-0.0cm}\\
\caption{\textit{Architecture of baseline and improved LID systems.}}
\label{fig:structures}
\end{figure*}

The deep architecture we employed is inspired by a recent system which performed extremely well for  LID~\cite{miao2019}, which used an front-end CNN to strengthen input feature extraction, followed by an LSTM to model temporal dependencies and learn long-range discriminative features over the input sequence.
We termed the final architecture the CLSTM system, from the concatenation of CNN and LSTM.
In addition to this architecture, which is applied here to DID for the first time, we make one further contribution in the ADI 2019 system described here, which is to introduce an implementation of the real-time iterative inversion (RTISI) method of speech time-scale modification (TSM) to augment the training data. The aim is to mitigate the effects of highly unbalanced data within the training set.
The method has two advantages, (i) it balances the dataset without using any other corpus (i.e. it is useful for challenges which disallow additional training data). (ii) RTISI, as a real-time algorithm, is efficient to implement.

The remainder of the paper is organized as follows: Section 2 describes DNN x-vector baselines and the improved structural changes leading to the proposed CLSTM architecture for DID; Section 3 introduces the TSM-based speed perturbation mechanism. Experimental results are presented in Section 4 while Section 5 concludes our work.

\section{language identification systems}
\label{sec:sec1}

\subsection{DNN x-vector}
The baseline end-to-end LID x-vector system based on a Time-Delay Neural Network TDNN~\cite{snyder2018spoken} structure is shown in Fig.~\ref{fig:structures}(a). 
Frame-level features centering on the current frame plus small context, are input to the first five layers. 
The following statistical pooling layer accumulates all frame-level outputs, calculates  mean and standard deviation to obtain a segment-level fixed-dimension representation. 
Segment-level statistics are then passed to two additional fully connected hidden layers and finally a softmax output layer.

\subsection{CLSTM}
The structure of the proposed CLSTM architecture is shown in Fig.~\ref{fig:structures}(b).
The front-end CNN learns how to extract local feature descriptors from input frames plus context. 
Directed by backward propagation assisted by an appropriate loss function, it automatically learns a temporal ordered feature representation, and acts like a sliding local feature extractor.

We also make use of LSTMs, since we believe that their powerful ability to model long-term dependencies, can be discriminative for languages.
Specifically, we add one LSTM layer between the TDNN layers to capture long-term variations in intermediate features, adopting an architecture~\cite{sak2014long} that has shown good performance in a related ASR task.

\section{Speech time-scale modification}
\label{sec:sec2}

\begin{figure}[tb]
    \centering
    \centerline{\includegraphics[width=0.9\linewidth]{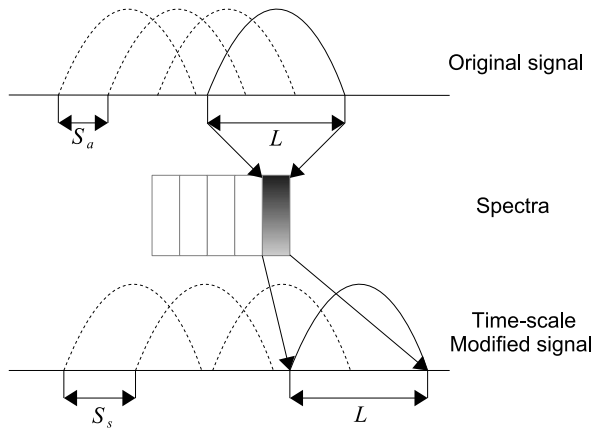}}
    \caption{\textit{Time-scale modification in RTISI.}}
    \label{fig:TSM}
  \end{figure}

The speech time-scale technique changes the speech rate by adjusting the duration of speech frames. 
This section introduces the theoretical basis of the real-time iterative inversion (RTISI) and its implementation. 
In recent years, several TSM algorithms~\cite{miao_expand_2018} have been proposed~\cite{chami2015architectural,chami2012real,dorran2003high,driedger2013improving}. 
This paper adopts the successful RTISI algorithm~\cite{beauregard2005efficient,zhu2007real}, which processes according to Fig.~\ref{fig:TSM}, in three stages,

1) It divides the speech signal $s$ into Hamming windowed frames, $x$ of length L and  window shift of $S_a$. 
\begin{eqnarray} \label{xx1}
x(\lambda)=s(\lambda.S_a:\lambda.S_a+L)h
\end{eqnarray}
where $h$ is the window function of length $L$ and $\lambda$ is the index of the frames.

2) The short-time Fourier transform magnitude (STFTM) is then obtained for each frame, 
\begin{eqnarray} \label{xx2}
\mathcal |X(\lambda, k) |= \bigg| \sum_{n=0}^{L-1} x(\lambda)e^{-j(2 \pi k n /L)} \bigg|
\end{eqnarray}
where $k$ represents frequency index. 

3)  The time-frequency domain signal $|X(\lambda,k)|$ is then transformed inversely back to the time domain with frame length $L$ and frame shift $S_s \neq S_a$.

The frame lengths  in the RTISI analysis and synthesis process are both $L=25$ms for our experiments.  We use a fixed synthesis step size $S_s = 10$ms.  If the analysis step size $S_a$ is smaller than the synthesis step size $S_s$, the synthesised speech rate is slower than the original speech, and the speech duration is longer than the original speech; conversely, if the analysis step size $S_a$ is larger than the synthesis step size $S_s$, the synthesised speech rate is faster than the original speech, and the speech duration is shorter than the original speech.  This is illustrated in Fig.~\ref{fig:TSM}.

The synthesis of each frame using RTISI only requires the previous signal, and not  the following signals, which ensures that the algorithm is structurally real-time and easy to implement.
We set a modification rate  $\alpha$ according to the analysis step size $S_a$ and the synthesise step size of $S_s$,
\begin{eqnarray} \label{xx3}
\mathcal \alpha = {{S_a} / {S_s}}
\end{eqnarray}

The relationship of speech duration between the original speech x  and the modified $ \widetilde{x}$ after the TSM transform is:
\begin{eqnarray} \label{xx4}
 length( \widetilde{x} ) = {{length(x)} / {\alpha}}
\end{eqnarray}

\section{Evaluation}
\label{sec:Evaluation}

\subsection{Database and experimental setup}
\label{ssec:Database&setup}
 
\subsubsection{Database}
\label{sec:Database}

The task of ADI is to perform dialect identification of speech collected from YouTube that belongs to one of the 17 target dialects. 
The training set comprises about 3,000 hours of Youtube recordings containing  Arabic dialect speech data from 17 Arabic countries.
The development and test sets comprise about 280 hours of speech data, similarly collected from YouTube. 
After automatic speaker linking and dialect labelling by human annotators, 57 hours of speech data were selected to use as development and test sets for performance evaluation. 
The test dataset is divided into three sub-categories on the basis of segment duration -- short (under 5\,s), medium (between 5 and 20\,s) and long duration (over 20\,s) utterances of dialectal speech.

\begin{figure}[tb]
    \centering
    \centerline{\includegraphics[width=1\linewidth]{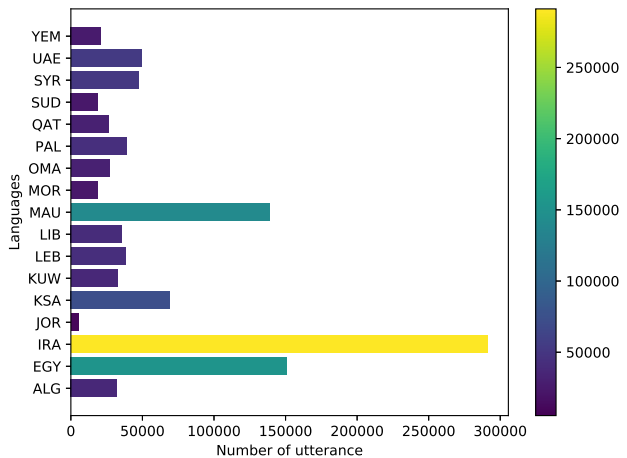}}
    \caption{\textit{Number of utterances for each of the 17 Arabic dialects in the training set.}}
    \label{fig:train}
  \end{figure}

To visualize this, Fig.~\ref{fig:train} plots a histogram of the number of utterances in the training dataset for each language. 
It is clear that the ADI training dataset is unbalanced mainly in two aspects; the large number of ``IRA'' utterances (and to a lesser extent ``MAU'' and ``EGY''), and the extremely limited number of ``JOR'' utterances. The latter, Jordanian dialect, has just 5514 training utterances.

\subsubsection{Experimental setup}
\label{sec: setup}

For the DNN x-vector and CLSTM architecture, the features are 23 dimensional MFCCs with a frame-length of 25ms, mean-normalized over a sliding window of up to 3 seconds. 
An energy-based speech activity detector, identical to that used in all systems, filters out non-speech frames. 
The DNN x-vector configuration follows~\cite{snyder2018spoken}.
The CLSTM architecture has two convolutional layers with 3$\times$3 filters followed by ReLU and batch normalization. 
The number of filters is set to 128 and 256 respectively, the cell dimension of the LSTM is 1024 and the recurrence of dimension is 256. 
For the time and frequency attention models, the number of hidden nodes is 64 and we use ReLU activation functions followed by batch normalization. All the experiments are carried out using Kaldi~\cite{povey2011kaldi}. 

It can be seen from Fig.~\ref{fig:train} that JOR has the minimum training data, therefore we choose to selectively augment the JOR data using TSM as described in Section~\ref{sec:sec2}. This specifically incorporates speeded up and slowed down versions of the same utterance. To do this, the synthesis step size $S_s$ is fixed to 10\,ms, the analysis step size $S_a$ is 8, 9, 11 and 12\,ms respectively, yielding a corresponding speech rate that varies from 0.8 to 1.2 times normal. Finally, these settings allow us to obtain a five-fold augmentation in the amount of training data (only for the JOR class, so the overall training time does not increase appreciably as it would if applied to every class). 

We also investigate using more traditional training data augmentation (TTDA) ~\cite{snyder2018spoken} for all languages to increase the amount and diversity of the existing training data, including additive noise from the MUSAN dataset, and reverberation (from the RIR dataset). These results are reported separately.

\begin{table}[tbh]
\caption{\textit{Performance results for DNN x-vector and CLSTM.}}
\begin{center}
\begin{tabular}{l | cc | cc}
\textbf{}  & \multicolumn{2}{c}{Dev} &  \multicolumn{2}{c}{Test} \\
\hline 
\textbf{System}  & \textbf{Accuracy}  & \textbf{EER}  &\textbf{Accuracy}  & \textbf{EER}  \\
\hline 
DNN x-vector &90.64 & 2.81 & 89.95 & 3.32\\
CLSTM  &91.20 & 2.55 & 90.82 & 2.91 \\
\end{tabular}
\end{center}
\label{table2}
\end{table}

\subsection{Experimental results}
 \subsubsection{DNN x-vector and CLSTM performance}

Table~\ref{table2} lists results from traditional DNN x-vector and the proposed CLSTM systems, separately listing accuracy and equal error rate (EER) for the development and test datasets. 
It is clear that the CLSTM architecture outperforms the DNN x-vector system; the combination of CNN and LSTM appears able to strengthen the feature extraction to learn directly from LID labels and model temporal dependencies to learn long-range discriminative language features over the input sequence.
This finding is similar to that obtained when the original architecture was applied to LID~\cite{miao2019}.

\subsubsection{CLSTM performance with TSM augmentation} 

JOR only has 5514 training utterances according to the analysis in Section~\ref{sec:Database}, therefore we use TSM as described in Section~\ref{sec:sec2} to augment JOR, eventually obtaining an expanded set of 27570  JOR training utterances ($5\times$).

Table~\ref{table3} compares the performance of the CLSTM with and without TSM used in this way. It reveals that adopting TSM for only the JOR data provides a benefit of around 10\% relative EER.

\begin{table}[htb]
\caption{\textit{Performance results for CLSTM with TSM.}}
\begin{center}
\begin{tabular}{ l | cc | cc}
 \textbf{}  & \multicolumn{2}{c}{Dev} &  \multicolumn{2}{c}{Test} \\
\hline 
 \textbf{TSM} & \textbf{Accuracy}  & \textbf{EER}  &\textbf{Accuracy}  & \textbf{EER}  \\
\hline 
No & 91.20 & 2.55 & 90.82 & 2.91\\
Yes & 92.26 & 2.38 & 91.07 & 2.68 \\
\end{tabular}
\end{center}
\label{table3}
\end{table}

\subsubsection{CLSTM architecture performance with TSM and TTDA augmentation applied} 
It was noted that we submitted the CLSTM architecture with TSM for the MGB-5 ADI challenge, because this system does not allow any other dataset (as incorporation of external data would violate the terms of the challenge).
However it is interesting to separately investigate the effect of TTDA, which uses additional noise and reverberation data.

Table~\ref{table4} shows that firstly applying either TSM on JOR or TTDA  on all languages are effective for both the development and test set accuracy.
The combination of TSM and TTDA  performed best for the test set, but interestingly was fractionally worse for the development set.

\begin{table}[htb]
\vspace{-1mm}
\caption{\textit{Performance results for CLSTM with TTDA.}}
\begin{center}
\small
\begin{tabular}{ l |  l |  cc | cc}
 \textbf{}  &  \textbf{}  & \multicolumn{2}{c}{Dev} &  \multicolumn{2}{c}{Test} \\
\hline 
 \textbf{TSM} &  \textbf{TTDA} & \textbf{Accuracy}  & \textbf{EER}  &\textbf{Accuracy}  & \textbf{EER}  \\
\hline 
Yes & No & 92.26 & 2.38 & 91.07 & 2.68 \\
No & Yes & 93.58 & 1.84 & 92.69 & 2.19 \\
Yes &Yes &  93.47 & 1.90 & 93.06 & 2.09 \\
\end{tabular}
\end{center}
\label{table4}
\vspace{-1mm}
\end{table}

  \begin{figure}[tb]
  \begin{minipage}[b]{1\linewidth}
    \centering
    \centerline{\includegraphics[width=1\linewidth]{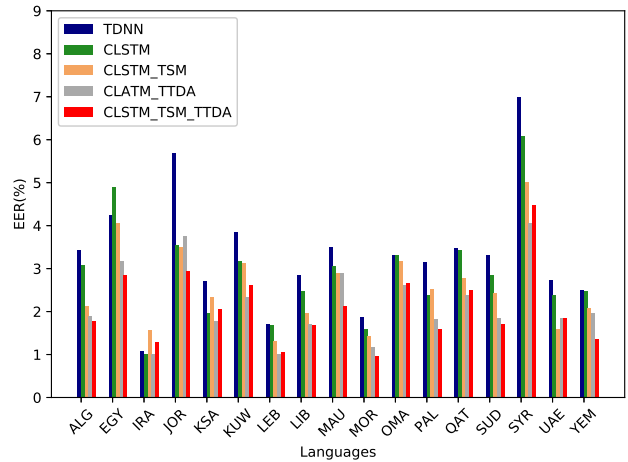}}
    \vspace{-2mm}
    \caption{\textit{EER of of individual systems for all dialects}}
    \label{fig:jor}
    \end{minipage}
  \end{figure}

\begin{figure*}[htb]
\centering
\includegraphics[height=8cm]{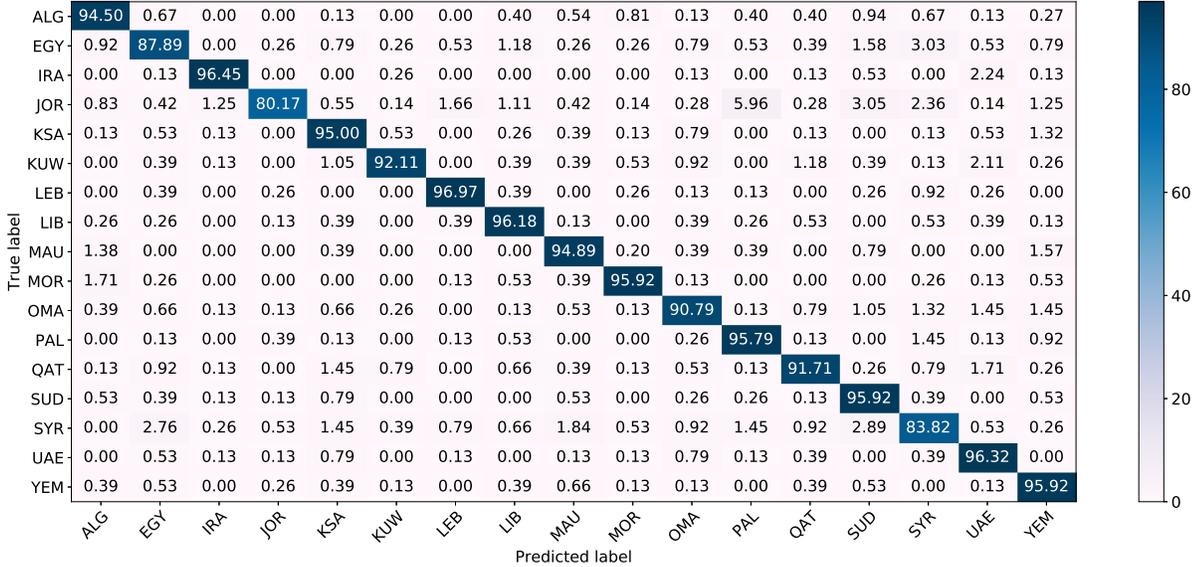}
\caption{\textit{DID confusion matrix (accuracy) for the CLSTM system, with both TSM and TTDA augmentation.}}
\label{fig:cm}
\end{figure*}

 \subsubsection{Further analysis}
To further explore the effect of different systems, we also plot  the EER achieved on the test set for each of the 17 Arabic Dialects in all the systems described above, in Fig.~\ref{fig:jor}.
We can observe that the EER for all dialects improves in line with the complexity of the architecture, and by a consistent degree for both TSM and TTDA -- which is unsurprising since the underlying task is very much  data-constrained (i.e. our architecture is powerful  but has insufficient data to unlock its potential).

Finally, we can explore the DID confusion matrix for all  dialects on the final CLSTM\_TSM\_TTDA architecture in Fig.~\ref{fig:cm}. 
Firstly, we can observe that the accuracy of most of dialects is well over 90\%. 
Lebanese (``LEB'') obtains the best performance of 96.97\% among all dialects.  It is noted that the smallest dialect ``JOR'', which is less well trained despite the augmentation, only achieves 80.17\%.
Interestingly,  both ``LEB'' and ``JOR'' belong to Levantine Arabic, i.e. they are in the same Arab dialect group. In this case, their similarity would be expected to lead to reduced performance (and increased confusion) between them. 
Moreover, the most three confused dialects with ``JOR'' are Palestinian (``PAL''), Syrian (``SYR'') and Lebanese (``LEB''). 
All of these similarly belong to Levantine Arabic, and share significant phonological, structural, and lexical features. 
However, at the same time, there are differences among Levantine dialects based on geographical division, and the grouping can be divided into North and South Levantine dialects. The Northern subgroup extends from Lebanon through Syria and into Turkey. 
Meanwhile South Levantine is spoken in Palestine, the western area of Jordan and in the south of Lebanon. 
The Palestinian dialect is reported to have slight phonetic differences from the North Levantine dialect. 
Meanwhile Arabic Syrian dialect is separately influenced by the Syriac language, a Semitic language of the Middle East which contains a large proportion of Arabic words but also significant numbers of loan words from Turkish and French. 
These differences appear to be useful in reducing the level of confusion.
It is uncertain whether these aspects are visible within the confusion matrix of Fig.~\ref{fig:cm}, yet they are clearly factors that affect at which level in the architecture the confusion occurs; this could range from mainly acoustic differences at the front end, through to structural, timing and even phonetic differences which are expected to be more confusing at the back end of the DID system. In future it would be extremely interesting to assess the sensitivity to confusion to identify where the main differences lie.

\section{Conclusion}
\label{sec:Conclusions}
This paper first  evaluated a new end-to-end LID architecture named CLSTM, employing a CNN front-end for a deep neural network structure with LSTM for extracting time sequencing information. 
Performance was shown to exceed that obtained by a traditional DNN x-vector architecture for a DID task.
Next we introduced a time-scale modification mechanism to increase the amount of training data available for the most constrained language in the unbalanced dataset, obtaining further improvements on the underlying proposed CLSTM architecture. 
We finally evaluated the effect of traditional training data augmentation for all the dialects, noting that the combination of the traditional training data augmentation and time-scale modification based on CLSTM architecture performed best.
The system described in this paper (excluding the TTDA augmentation which uses disallowed out-of-set data) was entered into the 2019 series ADI challenge (MGB-5), and achieved a 2nd place ranking out of the 15 MGB-5 ADI entries, as presented at ASRU\,2019.


\vfill
\pagebreak

\bibliographystyle{IEEEbib}
\bibliography{refs}

\end{document}